\documentclass[aip,jap,reprint]{revtex4-1}

\usepackage[utf8]{inputenc}
\usepackage{amsmath}
\usepackage{amssymb}
\usepackage{amsfonts}
\usepackage{graphicx}
\usepackage{siunitx}					
\usepackage{color}					
\usepackage {CJK}

\renewcommand{\d}{\mathrm{d}}

\begin{document}
\begin{CJK*}{UTF8}{}
\title{Dynamic stiffness of the contact between a carbon nanotube and a flat substrate in a peeling geometry} 
\CJKfamily{gbsn}
\author{Tianjun Li (李天军)}
\affiliation{Key Laboratory of Computer Vision and System (Ministry of Education), Tianjin University of Technology, Tianjin, 300384, P R China}

\author{Lor\`ene Champougny }
\affiliation{Univ Lyon, Ens de Lyon, Univ Claude Bernard, CNRS, Laboratoire de Physique, F-69342 Lyon, France}
\affiliation{Present address: ESPCI Paris, PSL Research University, CNRS, Gulliver,
IPGG, 6 rue Jean Calvin, 75005 Paris, France}

\author{Ludovic Bellon}
\email{ludovic.bellon@ens-lyon.fr}
\affiliation{Univ Lyon, Ens de Lyon, Univ Claude Bernard, CNRS, Laboratoire de Physique, F-69342 Lyon, France}

\date{\today}

\begin{abstract}
We study the physics of adhesion and the contact mechanics at the nanoscale with a peeling experiment of a carbon nanotube on a flat substrate. Using an interferometric atomic force microscope and an extended force modulation protocol, we investigate the frequency response of the stiffness of the nano-contact from DC to \SI{20}{kHz}. We show that this dynamic stiffness is only weakly frequency dependent, increasing by a factor 2 when the frequency grows by 3 orders of magnitude. Such behavior may be the signature of amorphous relaxations during the mechanical solicitation at the nano-scale.
\end{abstract}

\maketitle 

\end{CJK*}

\section{Introduction}

At the nanometer scale, the interaction between objects can be quite different from what expected at the macroscopic level: the large surface to volume ratio amplifies the effects of some phenomena, such as the Van der Waals forces. Adhesion is therefore ubiquitous, and impacts on the mechanical interaction between nano-objects. It has for instance a strong influence in atomic force microscopy (AFM)~\cite{Butt-2005}, where a nanometric tip is used to measure the topography by scanning a surface at constant interaction. It also drives the behavior of nano-particules, nano-wires or nano-tubes in contact with a substrate~\cite{Kis-2008}. Quantifying adhesion processes is thus important to understand the physics in the nano-world.

Atomic force microscopy, thanks to its ability to apply and measure forces and displacements in the nN and nm range respectively, is a first-choice tool to explore nano-mechanics. For example, the AFM has been successfully used to characterize the adhesion of one dimensional nano-object, such as carbon nanotubes~\cite{Hertel-1998-JPhysChemB,Hertel-1998-PRB,Kis-2006,Strus-2009-Nanotech}. In particular, in peeling experiments, one measures the force while the nanotube is pulled from a flat substrate\cite{Strus-2008,Strus-2009-CST,Strus-2009-PhysRevB,Ishikawa-2008,Ishikawa-2009,Xie-2010,Buchoux-2011,Li-2015}. The force versus distance curve displays a signature specific to this peeling process, and can be used to extract the energy of adhesion between the nano-object and the surface. Scanning electron microscopy has also been used to substantiate those conclusions by confirming the geometry of the contact during peeling~\cite{Ishikawa-2008,Ishikawa-2009,Ke-2010,Chen-2016}.

These peeling experiments are however restricted to the quasi-static mechanical behavior of the contact, since the time scales probed are of the order $\SI{1}{s}$. In this article, we explore the dynamic response of the contact, which can be quite different. Indeed, in previous works~\cite{Buchoux-2011,Li-2015}, we showed that at frequencies higher than $\SI{10}{kHz}$, the adhesion processes can be considered as frozen: the nanotube has no time to adhere to (or detach from) the substrate, leading to a dynamic stiffness which is greater than the static one. We are therefore interested in probing the intermediate frequency range: does any characteristic time scale govern this dynamic behavior~? What are the physical processes associated to this ``freezing'' of the adhesion~?
 
To explore this frequency range, we will extend one mode of operation of AFM, generally referred to as \emph{force modulation}~\cite{Radmacher-1993}. In this mode, one adds a small amplitude sinusoidal oscillation to modulate the force while the tip is in contact with the sample. Using an adequate model of the rheological behavior of the sample, the response of the AFM probe at this frequency leads to the ability of mapping rheological properties at the nanometer scale. In our case, such force modulation is tricky in two aspects. First, the amplitude of the modulation has to be extremely small to probe the linear response of the nanotube. Second, several frequencies should be measured at the same time, since the precise contact configuration can vary between successive experiments. We will introduce a protocol to address those two points, and measure the frequency behavior of the dynamic stiffness of the contact.

This article is organized as follows: in part \ref{section:quasistatic}, we first describe the experiment and the quasi-static force curve measurements, leading to the characterization of the adhesion energy between an individual carbon nanotube and a substrate of mica. In part \ref{section:dynamic}, we study the dynamic stiffness, either using a thermal noise analysis of the AFM probe during the peeling process, or performing extended force modulation experiments. In the last part, we discuss the experimental results and suggest some physical interpretation of the weak frequency dependence of the dynamic stiffness of the contact.

\section{Quasi-static force measurements} \label{section:quasistatic}

\begin{figure}
\centering
\includegraphics[width=85mm]{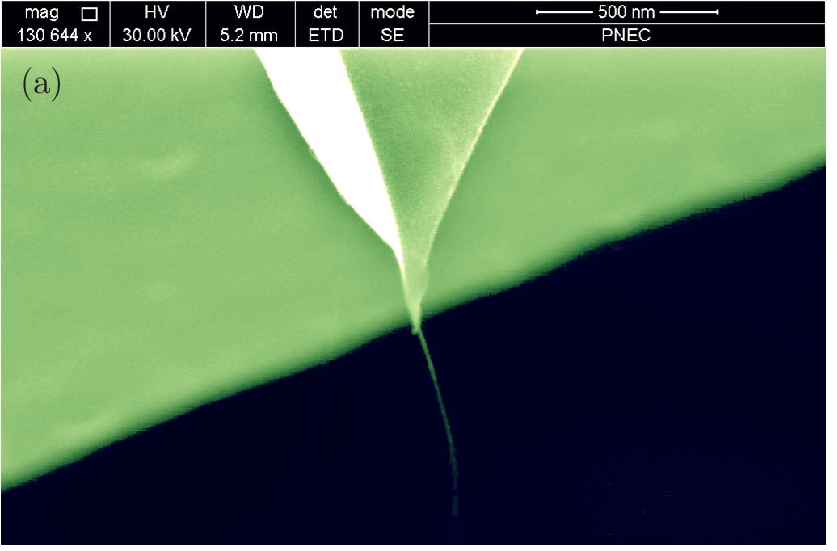}

\includegraphics{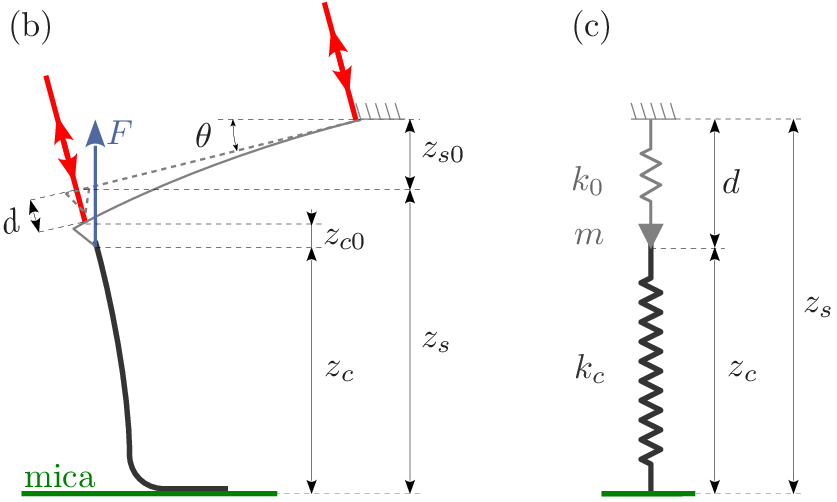}
\caption{(Color online) -- (a) Scanning electron micrograph of a soft CNT grown directly on an AFM tip. The nanotube length is $L_{c}\sim\SI{500}{nm}$. (b) When the nanotube is pressed almost perpendicularly against a mica surface ($\theta=\ang{15}$ inclination of the AFM cantilever with the substrate), part of the nanotube adheres to the surface due to Van der Waals interactions. The shape of the CNT is fixed by an equilibrium between the adhesion of the part in contact and the bending of the free standing part of the nanotube. From the measurement of the AFM cantilever deflection $d$ (using differential interferometry~\cite{Paolino-2013}) and sample position $z_{s}$, the force $F$ acting on the nanotube and its compression $z_{c}$ can be recorded. (c) The system is modeled by the effective mass $m$ of the cantilever being connected to two springs: the cantilever (spring constant $k_0$) attached to the static reference, and the nanotube in contact with the substrate (effective stiffness $k_c$). The sketch corresponds to the simplified equation where $\theta=\ang{0}$ and the origins are defined such as $z_{c0}=z_{s0}=0$.}
\label{fig:SEM-sketch}
\end{figure}

The nanotubes are grown directly~\cite{Marty-2006} at the tip apex of AFM probes by Chemical Vapor Deposition (CVD): the bare silicon cantilevers are fully dipped into the catalyst solution, then gently dried in a nitrogen flux before being placed in the furnace. CNTs grow everywhere on the cantilever, and around 1 in 3 cantilevers has a CNT at the tip~\cite{Li-2015}. A typical sample is shown by the SEM (Scanning Electron Microscopy) image of figure~\ref{fig:SEM-sketch}(a). Growth parameters are tuned to produce mainly single wall CNTs, but the effective sample may slightly differ from this goal. It may consist of a bundle of a few single wall nanotubes, carry a significant amount of amorphous carbon~\cite{An-2014}, or be a few-wall nanotube. No TEM (Transmission Electron Microscopy) images are available for our samples, since they tend to break during extensive testing, and thus no a posteriori imaging can be performed.

In the experiments, the nanotube is pushed against a flat mica substrate, as shown in the schematic diagram of figure~\ref{fig:SEM-sketch}(b). The translation of the substrate is performed with a piezo translation platform (Physik Instrumente --- PI P527.3) operated in closed loop, featuring an accuracy of \SI{0.3}{nm} rms thanks to its embedded capacitive sensor. We measure the deflection $d$ of the AFM cantilever with a home-made highly sensitive quadrature phase differential interferometer, which detects the optical path difference between the sensing beam (focused on the cantilever tip) and the reference beam (focused on the static base)~\cite{Schonenberger-1989,Bellon-2002-OptCom,Paolino-2013} --- see sketch in figure \ref{fig:SEM-sketch}(b). The deflection $d$ and the sample vertical position $z_s$ are simultaneously recorded with high resolution acquisition cards (National Instruments --- NI-PXI-446x) at $\SI{200}{kHz}$.

With both $z_s$ and $d$ being calibrated, using a proper definition of the origins ($z_{c0}$, $z_{s0}$) we can compute at any time the compression of the nanotube:
\begin{equation}
z_{c}=z_{s}-d\cos\theta
\end{equation}
where $\theta=\ang{15}$ accounts for the inclination of the cantilever with respect to the substrate. With the hypothesis that the horizontal forces acting on the nanotube are negligible, we can also compute the vertical force acting on the nanotube~\cite{Buchoux-2011,Li-2015}:
\begin{equation}
F=-\frac{k_{0}}{\cos\theta} d
\end{equation}
with $k_{0}$ the static stiffness of the AFM cantilever (calibrated from its thermal noise~\cite{Butt-1995}). Using compression instead of sample position allows us to take into account the compliance of the cantilever, thus to focus on the nanotube properties only in the force versus compression curves. In the following, we will drop $\cos\theta$ when writing the equations to ease their reading (thus taking $\theta=\ang{0}$), but taking into account this small correction is straightforward and has been done during the data analysis. Figure~\ref{fig:SEM-sketch}(c) reports the sketch the simplified model.

\begin{figure}
\includegraphics[width=0.5\textwidth]{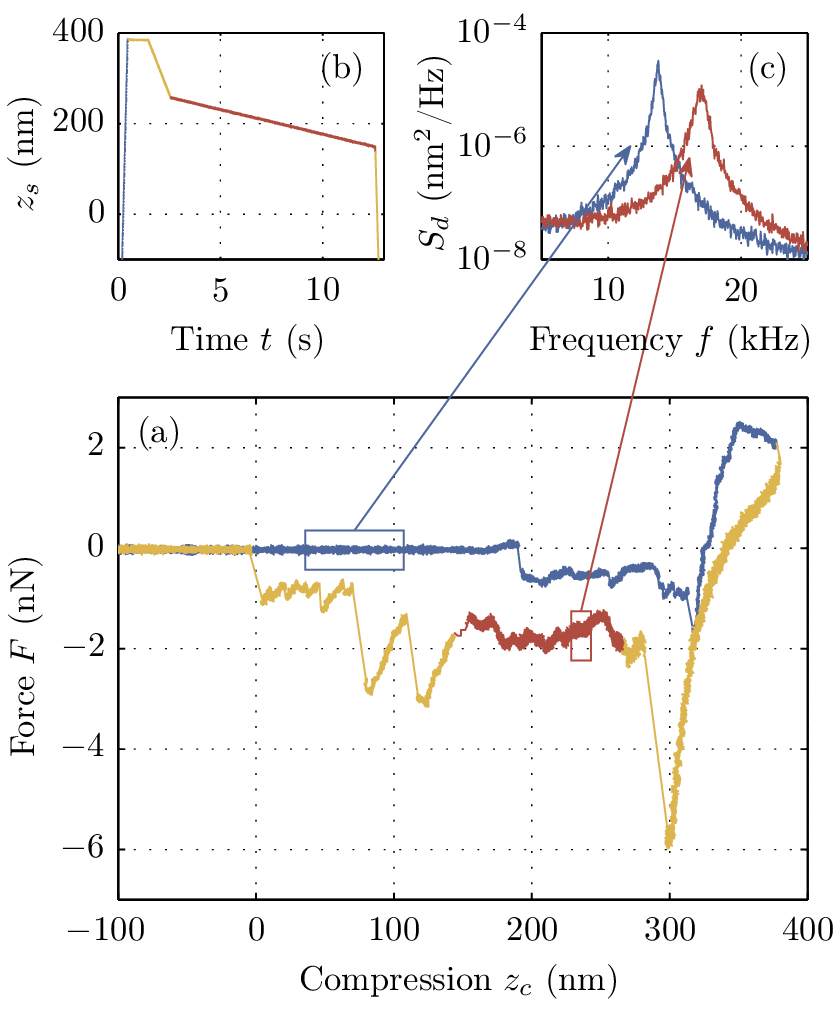}
\caption{(Color online) -- (a) Force $F$ of a CNT as a function of its compression $z_c$ on a mica substrate. A strong hysteresis, due to the adhesion can be noted between approach (blue) and retraction (yellow/red). Plateaus in the force curve are characteristic of a peeling process~\cite{Buchoux-2011}. (b) The ramp is designed to probe specifically the peeling configuration corresponding to compressions $z_{c}$ between $\SI{160}{nm}$ and $\SI{260}{nm}$ during retraction. (c) As shown by the PSD of the deflection $S_d$, the thermal noise of the cantilever allows the measurement of the additional stiffness due to the contact: the resonant frequency is higher (red) when the nanotube is in contact with the substrate than before contact (blue).}
\label{fig:forcecurve}
\end{figure}

An example of a force curve is plotted in figure~\ref{fig:forcecurve}(a). It presents a strong hysteresis due to the adhesion between approach and retraction, with a plateau-like behavior of the force during retraction: this is the signature of an adhesion and peeling process of the nanotube on the surface~\cite{Buchoux-2011,Li-2015}. The steep evolution at the largest compression corresponds to a hard contact between the AFM tip and the substrate, while the jumps between different plateaus during retraction are the signatures of defects in the nanotube. The shape of this force curve is robust to a few hundred cycles, showing an excellent reproducibility for different landing positions on the mica substrate. Other nanotubes present very similar features, except for the position of defects along their length which is naturally specific to each sample.

We focus our attention here on the longest force plateau during retraction of this nanotube, for compressions $z_{c}$ ranging between $\SI{160}{nm}$ and $\SI{260}{nm}$. We therefore use a non uniform approach and retraction speed to enhance the resolution in this area of interest, with a slow ramp at $\SI{0.1}{\micro m/s}$, as shown in figure~\ref{fig:forcecurve}(b). The force curve in this range of compression, though presenting some deviations to a perfect plateau, is rather flat: the local slope of the quasi-static force versus compression fluctuates around zero with a maximum slope below $\SI{3e-2}{N/m}$. The quasi-static stiffness of the nanotube in contact with the sample, defined by this slope, is thus close to zero in average: $k_{QS}=\d F/\d z_{c}= (4\pm22) \, \SI{}{mN/m}$ (mean and standard deviation).

\section{Contact dynamic stiffness} \label{section:dynamic}

As the compression ramp is sufficiently slow, we stay long enough around any compression $z_c$ to measure a spectrum of the deflection driven by the cantilever thermal noise or to probe the frequency response of the contact. The force acting on the AFM tip is no longer due to the deflection of the cantilever alone, since the nanotube touching the surface has to be considered as well. The mechanical oscillator (the first mode of cantilever) experience an effective stiffness $k_0+k_{c}$, shifting its resonance angular frequency\footnote{In this paper, we will equally use the natural frequency $f$ (in Hz) for the experimental data, and the angular frequency $\omega=2 \pi f$ (in $\SI{}{rad/s}$) for the equations.} from $\omega_{0}$ to $\omega_{c}$, as illustrated in figure~\ref{fig:forcecurve}(c)~\cite{Buchoux-2009,Buchoux-2011,Li-2015}. Dissipation is neglected in the following since the quality factor of the resonances is always large, above 20 at worse.

Let us model the system as sketched in figure \ref{fig:SEM-sketch}(c): the cantilever's dynamics is that of a simple harmonic oscillator (SHO, effective mass $m$, stiffness $k_{0}$), while the nanotube in contact with the substrate adds a stiffness $k_{c}$. The equation of motion of the tip can be written in the time space as:
\begin{equation}
m\ddot{d}=F_{\mathrm{ext}}-k_{0}d+k_{c}z_{c}
\label{eq:motion}
\end{equation}
where $F_{\mathrm{ext}}$ is an hypothetical external force acting on the tip. Moreover, the compression of the nanotube $z_{c}$ and the deflection of cantilever $d$ are linked by
\begin{equation}
z_{c}=z_{s}-d
\label{eq:zczs}
\end{equation}
Let us study two different cases, depending on whether $z_{s}$ is constant or externally driven.

\subsection{Quasi-static thermal noise measurement}

If $z_{s}$ is constant (the substrate is static --- or quasi-static), the equation of motion (\ref{eq:motion}) reads in the Fourier space:
\begin{equation}
\Big(k_{0} + k_{c}(\omega) - m \omega^{2}\Big)d(\omega)=F_{\mathrm{ext}}(\omega)
\end{equation}
In this last equation, we explicitly allow $k_{c}$ to depend on frequency, but as the resonance is sharp enough, only its value close to the resonance matters to accurately describe the thermal noise peak. This equation thus corresponds to a simple harmonic oscillator of mass $m$ and stiffness $k_{0} + k_{c}(\omega_{c})$. The SHO parameters (mass, stiffness) are linked to the resonance frequency by the usual relation:
\begin{align}
m\omega_{0}^{2} & = k_{0} \\
m\omega_{c}^{2} & = k_{0} + k_{c}(\omega_{c})
\end {align}
We can thus easily compute the contact stiffness $k_{c}$ from the angular frequency $\omega_{c}$ of the resonance~\cite{Buchoux-2009,Buchoux-2011,Li-2015}:
\begin{equation}
k_{c}(\omega_{c}) =k_0 \left[\left(\frac{\omega_{c}}{\omega_{0}}\right)^2-1\right] \label{eq:fshift}
\end{equation}

\begin{figure}
\includegraphics{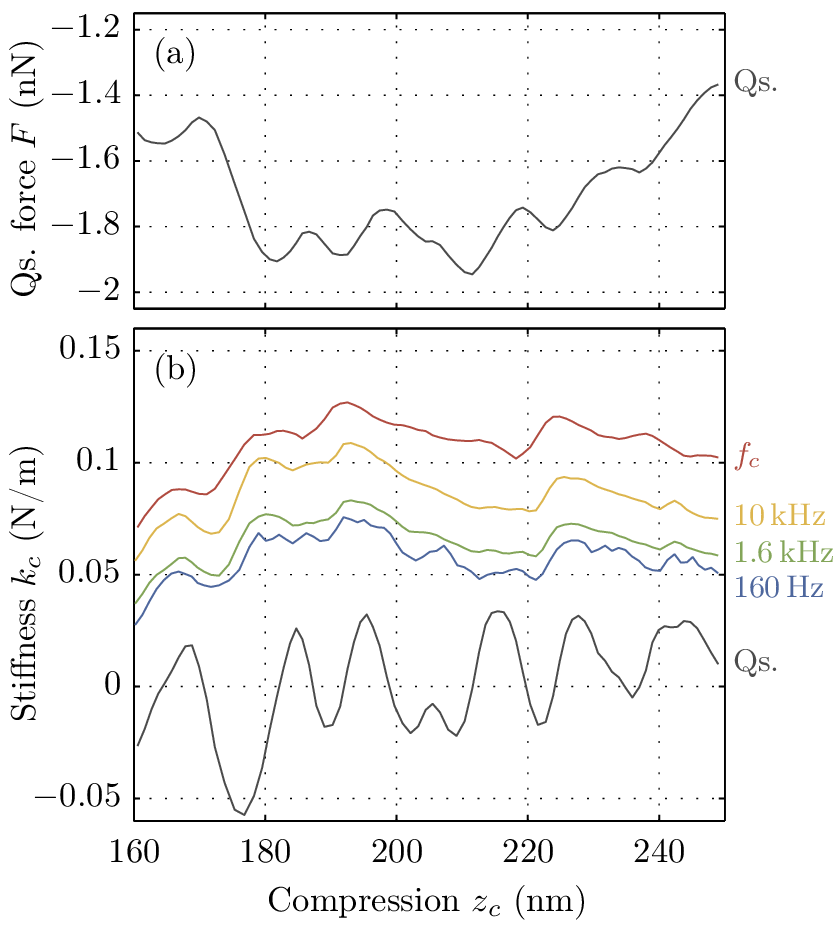}
\caption{(Color online) -- (a) Position dependence of the quasi-static force. (b) Contact stiffness: quasi-static (Qs.), dynamic from response measurement (at $\SI{160}{Hz}$, $\SI{1.6}{kHz}$ and $\SI{10}{kHz}$) and from thermal noise analysis ($f_{c}$). The dynamic stiffness is higher than the quasi-static one, and is almost constant on the force plateau. Some small variation are visible along the nanotube length though, they are correlated between force and stiffness: the larger the adhesion, the larger the dynamic stiffness.}
\label{Fig:allvsc}
\end{figure}

The resonance frequency is measured from the noise spectrum of the cantilever~\cite{Buchoux-2009,Buchoux-2011,Li-2015}: the power spectrum density (PSD) of the fluctuations of deflection, induced by the thermal noise random forcing, is sharply peaked at $\omega_{c}$, as illustrated in figure \ref{fig:forcecurve}(c). During the quasi-static ramp, we compute the PSD on a $\SI{0.4}{s}$ sliding time window, thus corresponding to only a $\SI{4}{nm}$ vertical translation of the sample. Using the procedure described above, we report figure \ref{Fig:allvsc}(b) the measured values of the stiffness $k_{c}(\omega_{c})$ of the contact as a function of the nanotube compression $z_{c}$. On the force plateau probed here, with this nanotube, we see that $k_{c}$ values at resonance are independent of the compression $z_c$, thus quite flat: $k_{c}(\omega_{c})=(108\pm12)\,\SI{}{mN/m}$ (mean and standard deviation). This result contrasts singularly with the quasi-static measurement: $k_{QS}= (4\pm22) \, \SI{}{mN/m}$.

\subsection{Dynamic measurements}

Thermal noise is a powerful tool to probe the contact dynamics, however it is limited to the resonant frequency of the micro-mechanical oscillator. To bridge the gap in the values of the contact stiffness between the quasi-static and the resonance conditions, we use an extended force modulation strategy: we shake the substrate vertically and test the response of the nanotube/substrate contact at various frequencies between $\SI{10}{Hz}$ and $\SI{10}{kHz}$. The substrate translation stage being too massive to reach such high frequencies, we mount the mica surface on an additional dedicated small piezo actuator (Physik Intrumente --- PICMA PL055, $\SI{5}{mm}\times\SI{5}{mm}\times\SI{2}{mm}$, with an unloaded resonance above $\SI{300}{kHz}$). The response function of this fast piezo to the driving voltage is calibrated using a stiff AFM probe in hard contact with the surface. It is flat at $\SI{16}{nm/V}$ in the frequency range probed here. From the driving voltage, we thus know precisely the induced displacement $z_{s}(\omega)$: our resolution is equivalent to that given by the interferometer, far beyond the sensor embedded in the translation stage. We can finally measure the response function $\chi(\omega)$ of the deflection $d(\omega)$ to $z_{s}(\omega)$. Using equations \ref{eq:motion} and \ref{eq:zczs} with $F_{\mathrm{ext}}=0$, we easily get 
\begin{equation}
\chi(\omega)=\frac{d(\omega)}{z_{s}(\omega)}=\frac{k_{c}(\omega)}{k_{0}(1-\omega^{2}/\omega_{0}^{2})+k_{c}(\omega)}
\label{chi}
\end{equation}
This equation is transformed to express the dynamic stiffness as a function of the measured response function $\chi(\omega)$:
\begin{equation}
k_{c}(\omega)=\frac{\chi(\omega)}{1-\chi(\omega)} k_{0}\left(1-\frac{\omega^{2}}{\omega_{0}^{2}}\right)
\label{eq:kc}
\end{equation}

\begin{figure}
\includegraphics{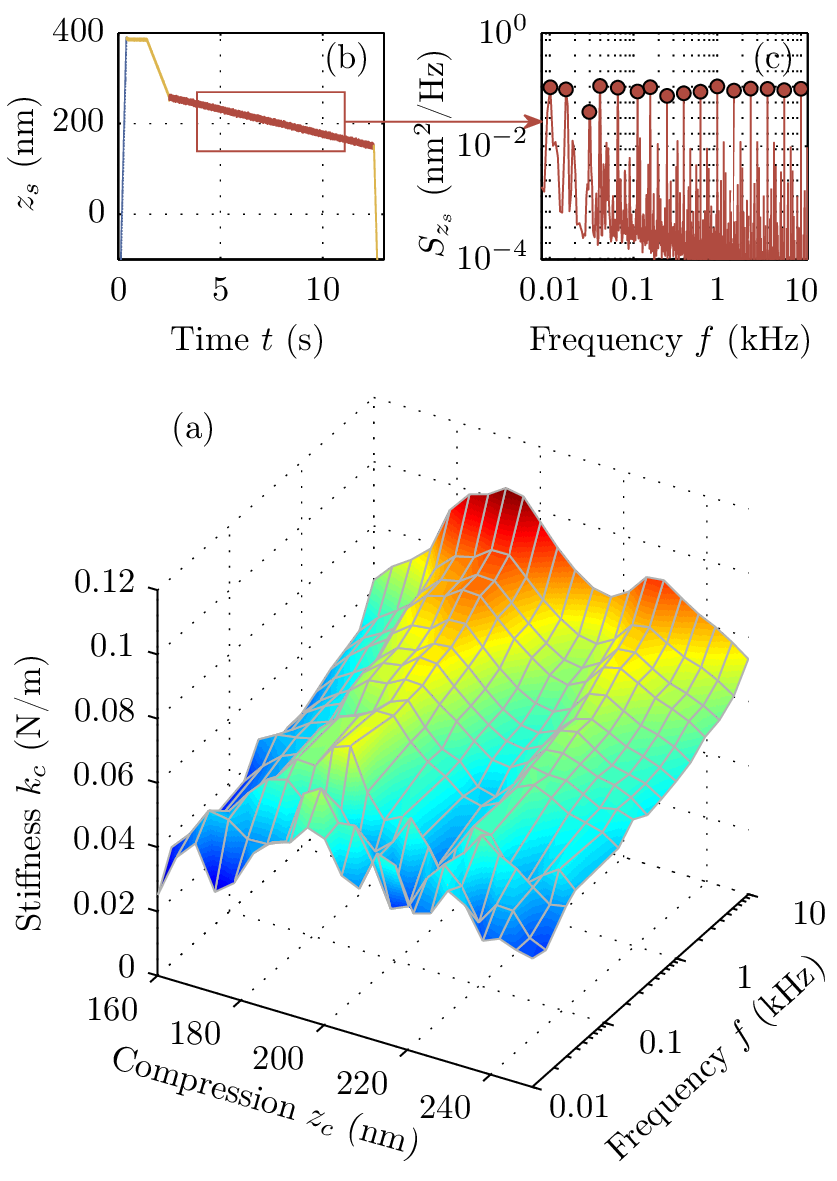}
\caption{(Color online) -- (a) Frequency and position dependence of the contact stiffness from the response measurement. (b) During the slow peeling ramp, a small amplitude vertical oscillation is applied to the sample: it consists in the sum of 16 sinusoids from $\SI{10}{Hz}$ to $\SI{10}{kHz}$ of $\SI{0.1}{nm}$ rms amplitude. (c) At the excitation frequencies (evidenced with filled circles on the PSD of the excitation), the response function $\chi(\omega)$ from this applied oscillation to the deflection of the cantilever is measured and translated into the equivalent contact stiffness. $k_{c}(\omega)$ depends slightly on the position along the nanotube, and is slowly increasing with frequency.}
\label{fig:kc3D}
\end{figure}

To evaluate $k_{c}(\omega)$ for different compression values of the nanotube during peeling, we add the periodic driving of $z_{s}$ on a ramp equivalent to that of used to test quasi-static and thermal noise properties of figure \ref{fig:forcecurve}(b). This periodic driving is designed as the sum of 16 sinusoids at frequencies approximately logarithmically spaced between $\SI{10}{Hz}$ to $\SI{10}{kHz}$, each having an amplitude of $\SI{0.1}{nm}$ rms (see figure \ref{fig:kc3D}). The response function $\chi(\omega)$ is then evaluated as the ratio of the Fourier transforms of the measured data ($d$ and $z_{s}$), at the driving frequencies only, each FFT being performed on a small $\SI{0.4}{s}$ sliding time window~\cite{Perez-Aparicio-2015}. This strategy allows one to measure the response function in the whole frequency range in a single approach-retract cycle, avoiding any question about the reproducibility of the precise peeling configuration encountered during successive cycles if one had to test only one frequency at a time. Frequencies smaller than $\SI{10}{Hz}$ were too noisy to be useful (they would require slower ramps to test each compression longer), and above $\SI{10}{kHz}$ the transfer function of the piezo is not flat enough to ensure precise measurements. Moreover by staying below $\SI{10}{kHz}$ we avoid any overlap with the natural thermal noise which is peaked around $\SI{17}{kHz}$ with a rms amplitude of $\SI{0.1}{nm}$, we thus keep a sufficient signal to noise ratio for each sinusoidal driving.

\begin{figure}
\includegraphics{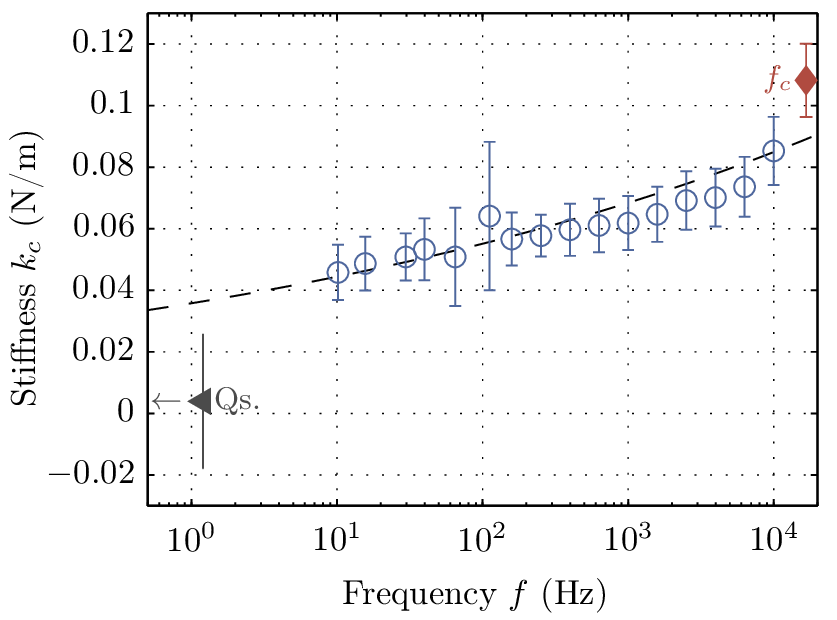}
\caption{(Color online) -- Frequency dependence of the average contact stiffness. Averaging the stiffness along the positions on the force plateau, we get this slowly increasing behavior of $k_{c}$ as a function of frequency (blue circles): it doubles when $f$ increases by 3 orders of magnitude. No characteristic time scale is evidenced on this curve, which can be fitted by a power law with a small exponent ($\propto f^{0.1}$, black dashed line). The high frequency behavior is compatible with the thermal noise estimation of $k_{c}$ at resonance (from the resonance frequency shift, red diamond). The low frequency behavior as well is compatible with the quasi-static estimation of the stiffness near 0 (gray triangle). Note that the quasi-static point should be placed at $f=0$, which cannot be seen on this logarithmic scale, so the reported point should be understood as the limit for $f\rightarrow 0$. Error bars correspond to one standard deviation in the average over position along the nanotube compression. Larger uncertainties around $\SI{100}{Hz}$ are due to a slightly higher environnemental noise in this frequency range, degrading the signal to noise ratio of the minute solicitation we are applying.}
\label{fig:kcvsf}
\end{figure}

The result of this process is plotted on figures \ref{Fig:allvsc} to \ref{fig:kcvsf}: $k_{c}$ as a function of nanotube compression for a few different frequencies in figure \ref{Fig:allvsc}(b), as a 3D plot versus compression and frequency in \ref{fig:kc3D}(a), and its mean value for every compression as a function of frequency in figure \ref{fig:kcvsf}. From these figures, we first observe that, at each frequency, the stiffness is roughly constant (the standard deviation is around $\SI{10}{\%}$ of the mean value). The small variations with compression are correlated at every frequencies and linked to the value of the adhesion force: the larger the adhesion, the larger the dynamic stiffness. The behavior with frequency is also quite weak: the stiffness grows steadily by a factor of 2 when the frequency spans 3 orders of magnitude. The value at high frequency tends to that predicted by the thermal noise analysis, and the low frequency trend is compatible with a zero stiffness for a quasi-static driving.

\section{Discussion and conclusion}

When the nanotube is close to the substrate, it tends to adhere and maximize its contact length. An equilibrium spatial shape balancing the adhesion energy of the adhered part and the bending energy of the free standing part is reached. For a long nanotube free of defect, this equilibrium translates into a force plateau in the force versus compression curve. This force plateau should thus correspond to a zero spring constant during the contact: the force doesn't depend on the position, thus no stiffness is expected. However, as evidenced by the thermal noise analysis, the resonance frequency of the AFM probe is higher during the peeling than when there is no contact. This behavior can be understood as adhesion and peeling being ``slow'' processes: at the resonant frequency (over $\SI{10}{kHz}$), if the contact between the nanotube and the substrate is seen as ``frozen'', then the free standing part of the nanotube is clamped on each side, and has a non zero stiffness. The question arising from this picture is then to understand what could impede the nanotube to relax quickly when it is retracted from or approached to the substrate, since the Van der Waals forces accounting for adhesion are an instantaneous interaction.

Using a direct excitation of the contact mechanics, we probe here its dynamic response. We find that the contact stiffness depends only weakly on frequency, as illustrated in figure \ref{fig:kcvsf}: it can be approximated by a power law in frequency with a small exponent ($\propto f^{0.1}$). In this peeling configuration, we thus do not find any characteristic time scale for slow vs. fast adhesion processes. Such amorphous behavior of the mechanics could be linked to relaxation processes with widely distributed time scales. One hypothesis is for example that nanoscale rearrangement or diffusion of defects occur during the strong mechanical solicitation of the contact point (the radius of curvature is typically only 10 times the nanotube diameter \cite{Buchoux-2011,Li-2015}). Such defects exist undoubtedly~\cite{Chen-2016}, as shown by the irregularities in the measured quantities along the cantilever length, in the form of fluctuations around the plateaus, or more dramatic force jumps in the force curves. Another hypothesis is that some friction occurs at the contact point, dissipating part of the mechanical energy and leading to delays in the system response. The last question is the role of the amorphous carbon unavoidably left around the nanotube during the growth~\cite{An-2014}, such as its plasticity during the mechanical solicitation and the friction that may occur at its interface with the nanotube or the substrate.

Answers to these questions may be found in the future by performing peeling experiments with simultaneous high precision force measurements and electronic microscopy visualisation of the nanotube configuration and of the contact shape at the nanometer scale. Comparison to experiments of peeling of other nano-objects (BN nanotubes, nano-wires) may also give interesting insights in the generality of the observed weak frequency dependence of the dynamic stiffness.

\begin{acknowledgments}
We thank F. Vittoz and F. Ropars for technical support, A. Petrosyan, S. Ciliberto, M. Geitner for stimulating discussions. We thank Anthony Ayari and the Plateforme Nanofils et Nanotubes Lyonnaise of the University Lyon 1 for the assistance in preparing the nanotubes. This work has been supported by the ANR project \emph{HiResAFM} (ANR-11-JS04-012-01) of the Agence Nationale de la Recherche in France. Finally, T. J. Li acknowledges the support from ZPNSFC (No. LQ16A020003), NFSC (No. 11547120), CPSF (147899), and the Scientific Research Foundation for the Returned Overseas Chinese Scholars (MOE of P. R. C.).
\end{acknowledgments}

%

\end{document}